# Aspects of Assembly and Cascaded Aspects of Assembly: Logical and Temporal Properties

Nicolas FERRY[1,2], Jean-Yves TIGLI[1], Stéphane LAVIROTTE[1], Gaëtan REY[1] and Michel RIVEILL[1]

[1] I3S, University of Nice – Sophia Antipolis
Sophia Antipolis, France

[2] CSTB (Centre Scientifique et Technique du Bâtiment)
Sophia Antipolis, France

**Abstract**
Highly dynamic computing environments, like ubiquitous and pervasive computing environments, require frequent adaptation of applications. This has to be done in a timely fashion, and the adaptation process must be as fast as possible and mastered. Moreover the adaptation process has to ensure a consistent result when finished whereas adaptations to be implemented cannot be anticipated at design time. In this paper we present our mechanism for self-adaptation based on t he aspect oriented programming paradigm called Aspect of Assembly (AAs). Using AAs: (1) the adaptations process is fast and its duration is mastered; (2) adaptations' entities are independent of each other thanks to the weaver logical merging mechanism; and (3) the high variability of the software infrastructure can be managed using a mono or multi-cycle weaving approach.

**Keywords:** *Aspect oriented programming, Context-awareness, Dynamic adaptation, Component Based Software Engineering.*

## 1. Introduction

**Background and motivation:** Ubiquitous computing relies on processing units present everywhere, at any times and in any things. The software infrastructure, on which a ubiquitous system is based, appears to be dynamically populated by the functionalities of such devices. Indeed, these services, potentially numerous, heterogeneous and m obile, may appear or disappear into it. These three characteristics (multiplicity, heterogeneity and mobility) induce the high variability of this infrastructure and therefore of ubiquitous systems. They must be adapted to this infrastructure and the adaptation mechanism must be able to manage this variability. Moreover, because of devices mobility, it is not possible to predict *a priori* which adaptations will be applied, but also how they should be composed. And all this must be achieved whilst maintaining reasonable and mastered response times.

**The problems:** In this paper, we address the issue of ensuring the continuous and dynamic adaptation of an application to changes occurring in its infrastructure (also called operational context), whilst considering the unpredictability and variability of this infrastructure, in a timely fashion, with mastered response time. Unlike approaches in which all the configurations or all the various compositions of adaptations are anticipated (and then bounded) at design time [1,2], we want to bring out (emergence) applications [3] according to their infrastructure in an unanticipated [4] manner. Thus, adaptations have to be independent of each other and the adaptation mechanism must be able to compose them, whilst ensuring the consistency of the resulting application. The variability that must manage the adaptation mechanism spreads on two axes: (1) on the devices available for a configuration described in an adaptation and (2) on the adaptations to compose. An adaptation entity does not have to be aware of others in order to be composed with them, ensuring a good separation of concerns and facilitating the evolution of adaptation concerns.

Such adaptations should be made whilst considering the dynamics of the changing infrastructure, to ensure that stable and usable applications are maintained. Adaptation response time is a major challenge for ubiquitous systems. As highlighted in [5], a ubiquitous system must not be too slow in reacting to changes, and should, for example, not use a service that is no longer present in its infrastructure. Moreover, the adaptation period should be sufficiently short to ensure that the system is not unavailable, or partially unavailable, for unacceptably long periods of time. However, response time is often ignored by projects requiring complex context processing, such as ontologies, for which execution time is unbounded [6], sometimes requiring several seconds [7].

**Our solution:** We have seen that in the field of ubiquitous computing, adaptation should be dynamic. In order to manage the heterogeneity of the devices included in the infrastructure of an application, we rely on service-oriented middleware [8], providing mechanisms to monitor it. Our mechanism for self-adaptation is primarily dedicated to service-oriented middleware whose services are orchestrated using component assemblies [9, 10]. These middleware also provide a range of services to manage the appearances and disappearances of services, which are directly implemented in the appearance and disappearance of components in the platform [11]. As we can see in the literature [12,13,14], compositional adaptation [15] is well suited to handle infrastructural changes. The loose-coupling between components facilitates their dynamic replacement, which makes them a particularly suitable approach for adaptive systems using compositional adaptation [16, 17].

As highlighted in [2], adaptation logic and application business logic have to be clearly separated. Moreover, since we do not want to anticipate the adaptations, they must be encapsulated into entities independent of each other. It allows them to be deployed without *a priori* knowledge of other adaptations. In order to achieve such adaptations, we propose an original approach based on aspect-oriented programming (AOP) [18], called "Aspect of Assembly" (AA). AOP is a way to achieve separation of concerns (SoC). Dynamic aspects allow adapting an application at runtime whilst encapsulating the adaptation into aspects [19]. Thanks to this encapsulation, the modularity of adaptations is improved and they can be more easily reused. However, classical AOP approach still suffers limitation in term of software evolution because interference management at runtime needs to be anticipated [20]. AA is a mechanism for the self-adaptation of an application to changes occurring in its infrastructure. Adaptations are in the form of compositional adaptation of components assembly with short and mastered response times. The adaptation process can involve one (mono-cycle) or several (multi-cycle) weaving operations (Fig. 1). Their composition does not require to be explicitly managed, and thus an AA can be deployed without considering others AAs.

**Case study:** Throughout the paper we will use the following scenario to illustrate these concepts. This scenario takes place in the context of a hospital. *The hospital, for ecological reasons, decided to implement a policy to reduce its energy consumption. Eve is a nurse at the hospital, when she enters a r oom the system would enable the switch to open the shutters rather than turning on the lights when the outside brightness is sufficient. She is entering in the room 500, newly assigned to an ol d woman who is visually impaired. The old woman's profile is a pr iority when entering a r oom, so in such a c ase artificial lighting is always used.* In section 3.3 a more complex scenario, used in the French ANR project called "Continuum" will also be used to illustrate our work in terms of response times.

**Outline:** The remainder of our paper is organized as follows: first we will present AAs, their mono-cycle weaving and our approach to manage interferences between AAs in an unanticipated way. In the following section, we will present their multi-cycle (Fig. 1) weaving and explain how it can preserve the same properties as the mono-cycle approach. Afterwards we will conduct a performance evaluation of the approach and show that adaptations' times are mastered. Finally we will study some related works before concluding.

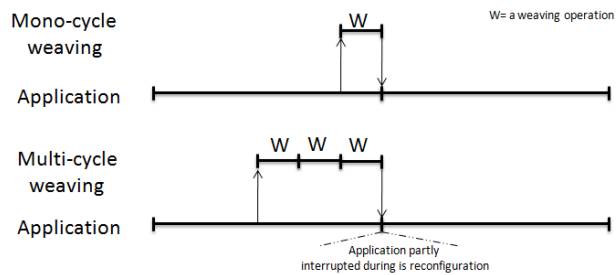

Fig. 1 Mono and multi-cycle weaving timelines

## 2. Aspect of Assembly

Aspect of Assembly (AA) is a model based on AOP for adaptation schemas. They allow structural reconfiguration of components assemblies at runtime, keeping black-box property of components. Modifications they induce are thus based on adding components and bindings between them. In traditional AOP, aspects are composed of pointcuts and advices. Pointcuts point out "where" to inject the code to weave while advices describe the code to be injected thanks to the aspect weaver. Pointcut genericity allows an aspect to be woven in many parts of the application. Thus, AOP minimizes code dispersion, grouping it in to reusable entities. Joinpoints represent all hooks of applications where advices can be woven. Classically, aspect languages provide mechanisms to add behavior to pointcuts thanks to operators *after*, *before* and *around* [18]. In the context of AA, these concepts are still valid but with some deviations. An advice describes a structural reconfiguration of a co mponents' assembly, while a pointcut identifies components' ports on which changes will take place. Thus, joinpoints are all entities of the assembly that structurally represent the application, on which changes will take place: components and their ports. The result of the weaving of AA is a s et of basic instructions such as adding a link or a component. Thus, our approach can be applied to several types of dynamic components platforms like SCA [10] or SLCA [9], for instances.

**Pointcuts** are defined as sets of filters on joinpoints meta-data (port ID or name, port type). Those filters construct lists of parameters satisfying the list of variables of the associated advice. They are the set of components ports on which the advice will be woven. For each generated list including a joinpoint for each pointcut variable, the advice is duplicated and the variables are syntactically replaced in the advice to match the base assembly joinpoints. Thanks to pointcuts, AAs are applied on components assemblies which are not necessarily known *a priori*. Pointcut are a way to manage the variability of the software infrastructure, thanks to duplication to manage homogeneous crosscuts [21], and to wildcards and metadatas to manage heterogeneous crosscuts [21]. For our experiments, we choose for convenience to express filters using some simple pattern matching as regular expressions on components, ports name and meta-data, and meta-data evaluation. As an example, the pointcut from the AA presented in Figure 2 describes that the variable `Shutter` will be associated to all pairs *composant.port* whose names is beginning by shutter with a port *SetState*. Line 3 associates the variable `light` components whose type is light and with energy consumption under 50W.

**Advice** is not a piece of code to be woven into the application's base code, but a set of component instances and links that will be woven inside an assembly of components. They can be considered as component assembly factories. To do this, advices are composed of a set of rules. These rules define which components or bindings between components have to be instantiated. An advice describes a set of adaptation rules to be applied on variable components defined in pointcut. Advices are basically specified in a DSL that we will present in the next section. We will describe later in section 2.3 how this language can be extended with a well-defined set of composable operators. Their merging with each other will be well-defined and to provide some properties in order to compose adaptations in an unanticipated way.

2.1 A language for Aspect of Assembly advices

Table.1 Advice language keywords

|  | Keywords / Operators | Description |
|---|---|---|
| **Port types** | comp.port | A provided port. |
|  | comp.^port | A required port. |
| **Rules for structural adaptations** | comp : type | To create a black-box component |
|  | comp : type(prop=val) | To create a black-box component and to initialize properties |
|  | required_port -> (required_port) | To create a link between two ports. |
|  | provided_port -> (required_port) | To rewrite an existing link by changing the destination port |

Advices are based on three types of rules: (1) the addition of black-box components, (2) rewriting links between components of the assembly and (3) the creation of new links. Rewriting involves components ports, it consists in forwarding an input port or redirecting a message (output port). These rules are identified thanks to two key words, ':' for black-box components instantiation and '->' for rewriting and creating links.

Figure 2 presents an example of AA written using the basic language defined in Table 1. We define an independent adaptation schema for our scenario. Another AA is required to achieve the scenario; it will be described later (Figure 5). It aims to link a switch and an RFid reader to a decision component which is bound to the shutter and the light. When an ID is read, the decision component checks if the ID is valid and if no visually impaired person is in the room then allows the user to turn on the light and automatically close the shutter (or inversely). Let's consider that some proxy components to communicate with the light, switch, shutter and RFid are generated and instantiated into the component assembly. We will now study the code of this aspect. It is called *IdentityManagement*. The four variables *Shutter*, *RFid*, *light* and *switch* associated to the name of the advice describe the joinpoints, identified by the pointcut matching that will be used in the advice. At lines 2 and 3 some black-box components are added. The rules described at lines 6, 8, 10, 12 and 14 define five new links. For exemple, line 12 aims to link the required port *DecisionEntity.^ShutterManagementEvent* to a port associated to the variable Shutter, for instance S*hutter.SetState*.

```
Pointcut:
1 Shutter:=/Shutter*.SetState/
2 RFid:=/rfid.*/
3 light:=/*(@type=light&energyConsumption < 50).*/
4 switch=/switch.^value_Evented_NewValue/
Advice :
1 schema IdentityManagement(Shutter,RFid,light,switch):
2 Decision : 'WComp.BasicBeans.DecisionEntity';
3 Timer : 'WComp.BasicBeans.Timer';
5
6 Timer.^Status _New_Evented_Value -> (Decision.SetTime)
7
8 Rfid.^ value_Evented_NewValue->(DecisionEntity.Manage)
9
10 switch ->(DecisionEntity.Manage)
11
12 DecisionEntity.^ ShutterManagementEvent->(Shutter)
13
14 DecisionEntity.^ LightManagementEvent->(Light.SetState)
```

Fig. 2 IdentityManagement Aspect of Assembly

## 2.2 Mono-cycle weaving

In the manner of automaton cycles consisting of successive phases of (1) data acquisition, (2) processing and ultimately (3) production of outputs, we speak of **weaving cycles**. For a cycle, weaver's input are: an assembly (the original application), called the base assembly, and a set of AAs. As a result, the weaver produces a final assembly (the adapted application). Figure 3 presents the weaving cycle involving the two AA of our scenario. Because the base assembly is composed of the five required components, both AAs are woven. A weaving cycle can be triggered on the appearance or disappearance of a component in the assembly or when they are selected or unselected. Each weaving cycle is processed on the base assembly free of any AA adaptation. The number of type of configurations of the system that can be described in a weaving cycle is equal to $2^{card(A_n)}$ where $A_n$ is a set of AAs. This means that the number of configurations described is $2^{card(A_n) \times (1+p_d)}$ where $p_d$ is the probability of having AAs duplicated. The weaving process can be formally written as: $T(Ass_0; A_n) = Ass_{n+1}$ where $Ass_0$ is the base assembly. This means that without using AA $2^{card(A_n) \times (1+p_d)}$ assemblies would have been designed to provide the same variability to the system.

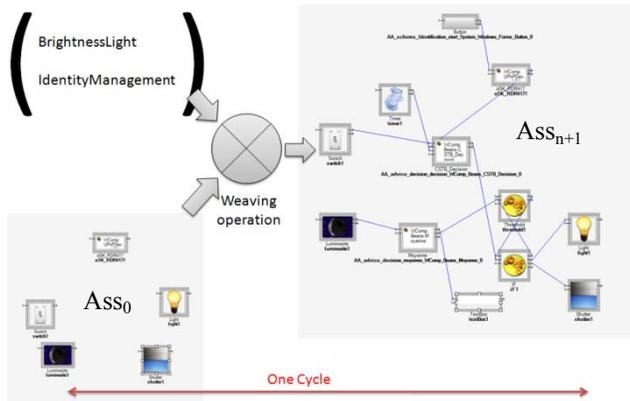

Fig. 3 Mono-cycle weaving

The weaving process can be divided into 5 majors steps (Fig. 4). First, pointcut matching is a function that takes a set of joinpoints from the base assembly and pointcuts, from a set of selected AA, as input. Its goal is to find the joinpoints on which advices will be woven… The second step aims to generate several combination of the joinpoints obtained during the pointcut matching. Each combination of joinpoints is composed of a joinpoint for each pointcut rule. The third step is called the advice factory. It generates instances of advices, replacing variable components in advices of selected aspects by joinpoints from combination obtained during the second step. Instances of advices describe modifications to be woven in the base assembly of components. Based on pointcut matching and joinpoint combination results, an advice can be woven several times during the same weaving process. These three first processes of the weaving mechanism are duplicated for each AA processed. Meaning that for each AA and for each process an algorithm can be selected. Finally, the composition engine merges all instances of advices with the initial assembly. It generates a single instance of advice that will be woven as the final assembly.

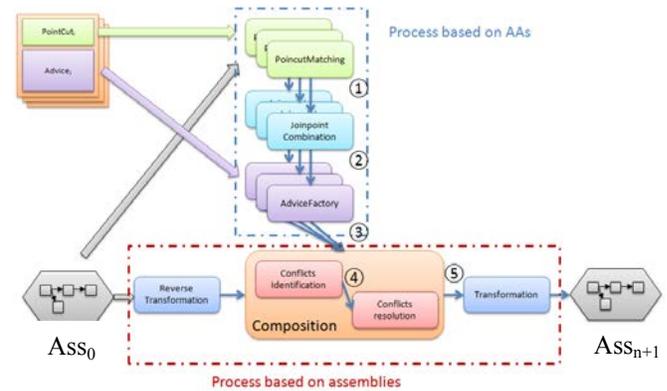

Fig. 4 The weaving process

## 2.3 Mono-cycle merging

The composition process can introduce interferences between AAs, between advices' rules. Interference is defined as: *"a conflicting situation where one aspect that works correctly in isolation does not work correctly anymore when it is composed with other aspects."* [22]. Various techniques exist to manage these interferences as the precedence between aspects that can be found in classical AOP [23] or the use of contracts as in [16]. They add some global predicates to aspects that an aspect can use to refer to another aspect, limiting the separation of concerns.

Our approach is to merge rules that interfere with each other and not to prevent explicitly interferences. It allows having AAs independent of each other that can be composed in an unanticipated manner and that can be easily added or removed by various actors. For this composition to be deterministic, meaning that the resolution of interference produces the same result, regardless of the order in which AAs are woven, it is necessary for the composition operation, as for the weaving operation, to be symmetrical. This symmetry property itself consists of three sub-properties: associativity, commutativity and idempotency. These properties: (1) allow the weaving process to be deterministic, (2) ensure that the order in which AAs are woven does not matter, (3) ensure that the system is confluent (because deterministic and symmetric) and (4) terminal (thanks to idempotency). In order to respect this property, it is necessary that:

- A pointcut cannot express negatives pointcut rules (i.e. rule requiring the absence of a component) (This may lead to the loss of the commutativity property).
- A pointcut cannot match components instantiated by another AA. (This may lead to the loss of commutativity and associativity properties).
- An advice cannot suppress components or bindings explicitly (This may lead to the loss of associativity and commutativity properties). Components or binding are suppressed if the AA is withdrawn.
- The rules' composition operation is symmetrical

Within these constraints, the only possible interference between AAs appears when a single joinpoint is used in several advices' rules. Those joinpoints are called shared joinpoints.

To enable the merging of these interfering rules with the previous properties, we constrain the advice language. Whatever the language used to write the advices, it must be based on a limited set of operators with a well-known semantic that can be merged. To be symmetrical, the merging operation of advices' rules requires that the merging operation of these operators is symmetrical. This property must be ensured between all operators. Adding an operator will require demonstrating that its merging with any other operator is symmetric. Those operators do not necessarily need to be themselves symmetrical.

For example, we defined the ISL4WComp language [24] as an extension of the previously defined DSL. ISL4Wcomp is based on the ISL Interaction Specification Language that describes patterns of interactions between independent objects [25]. ISL4Wcomp adapts these specifications to consider interactions based on messages or events between components. In this language, 6 operators were defined; they are presented in Table 2.

Table.2 ISL4WComp operators

| Operators | … ; … | sequence |
|---|---|---|
| | … ‖ … | parallelism |
| | If(condition) {…}else{…} | Condition is evaluated by a blackbox component |
| | Nop | Nothing to do |
| | Call | Allow to reuse the left part of a rule in a rewriting rule |
| | Delegate | Allow to specify that an interaction is unique in case of conflict |

As an example, the aspect presented in Figure 5 proposes to adapt the behavior of the AA described in Figure 2 by adding an energy saving concern as described in the scenario. To be applied it requires a brightness sensor, so that the user can turn on the light only when the brightness is under a defined threshold. Moreover, the new assembly opens the shutter when the user tries to switch on the light while the brightness is too high. We will now study the advice's code of this AA. It is called *brightness_light*. The three variables *light*, *brightness*, *shutter* associated to the name of the advice describe the joinpoints, identified thanks to the pointcut matching, that will be used in the advice. This AA highlights the three types of rules previously defined. At lines 3 and 4 some black-box components are added. The threshold component is instantiated with the property threshold up to 10. A property is a public variable from a component available through its interface. Lines 6-10 define an input port rewriting rule. All links connected to the input port (method) *SetState* will be rewritten. This rule involves the operator if, this mean that a *if* component will be instantiated. The condition to be evaluated by this component comes from a call on the method *IsReached* from the *threshold* black-box component. If the condition is true, then the shutter is open, else the rewritten link is done. Rules defined at lines 11, 12 allow defining two new connections. As an example, the second rule links the output *NewAverage* from the black-box component *Average* to the input method *SetValue* from the black-box component *threshold*.

```
Pointcut
1 light:=/light[[:digit:]].SetState/
2 Shutter:=/shutter[[:digit:]].SetState/
3 Brightness:=/brightness*.*/
Advice:
1 schema brightness_light ( light, brightness, switch ) :
2
3 threshold : 'BasicBeans.Threshold' ( threshold = 10 )
4 Average : 'WComp.BasicBeans.Average';
5
6 light -> (
7    if (threshold.IsReached)
8         {Shutter }
9    else
10        {call})
11 Brightness.NewValue -> (Average.AValue)
12 Average.NewAverage -> (threshold.SetValue)
```

Fig. 5 Brightness_Light Aspect of Assembly

Thus, the composition mechanism embeds a merging mechanism based on theses operators. Conflicting rules are expressed in the form of trees. Operators are the nodes of these trees and port their leaves. Merging two trees consist in merging the operators according to pre-defined rules. The 24 rules are defined in [24]. The merging of each of these operators has been defined as symmetric in [24]. The merging operation of two operators can be described with several rules. As an example, the merging operation of two *if* operator is based on two rules. Lets write $\otimes$ as the merging operation, *if(condition1,thenA,elseB)* $\otimes$ *if(condition2,thenC,elseD)* is equal to:

- If condition1 = condition2 :
  *if(condition1,thenA $\otimes$ thenC, elseB $\otimes$ elseD)*
- If condition1 ≠ condition2:
  *if(condition1,if(condition2,thenA $\otimes$ thenC, thenA $\otimes$ elseD), if(condition2, elseB $\otimes$ thenC, elseB $\otimes$ elseD))*

The merging operation is then propagated to the leaves. When two rules adding two bindings do not use operators and are conflicting, the result of the merging operation consists in adding a *parallel* operator between the two bindings. This also ensures the symmetry property of the merging operation. Finally, a rule adding a black-box component cannot cause a conflict since an AA cannot reuse a component instantiated by another AA. Once the trees are merged, they are transformed into elementary instructions (add/remove component/binding), operators are then represented in the assembly by components with a well-known semantic.

As an example, when both AAs presented in section 2.1 are composed, a conflict occurs on the port *switch.^on*. The result of the merging operation of the two conflicting rules is described in Figure 6.

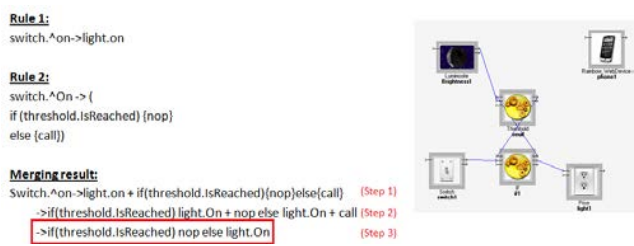

Fig. 6 An example of ISL4WComp rules merging

In this example we can see that the merging operation is propagated to the leaves. First it merges the message *light.on* and the operator *if* (step 1). Then it merges the message *light.on* and the *nop* operator in the *then* branch of the *if* and the message *light.on* and the *call* operator in the *else* branch (step 2). Because *nop* is an absorbing operator the result for the *then* branch is *nop*. Conversely, *call* is a neutral operator so the result for the *else* branch is *light.on* (step 3).

### 2.4. Multi-cycle weaving

AA's weaver also allows to chain several rounds of weaving so that an adaptation can be described using (and be the result of) several weaving cycles. Thanks to this multi-cycle approach, we will be able to decompose AAs according to their functional production and to reuse a functionality already woven.

Introducing this decomposition provides facilities for the reuse of parts of an AA. It also improves its evolving facilities. This means that it will be easier to identify which part of the system remove or swap according to the context. As an example, in our scenario, according to the rooms visited by the nurse, the mechanisms to monitor the brightness can change; it may be a sensor into the room or a weather service of the hospital for not equipped rooms. To make such changes, we must clearly identify the functional production of an AA in order to know which AA need to be exchanged and not to group all these productions in a single AA. The latter would imply rewriting the whole AA for each configuration. However, an AA cannot reuse a component instantiated by others AAs.

Therefore, the multi-cycle approach proposes to group AAs according to the functionality they intend to weave and to dedicate a functional group to a weaving cycle. Classically, for a ubiquitous application, we will create three groups and therefore three cycles of weaving: a cycle for a group of AAs that produces the perception mechanism, a cycle for a group of AAs that produces the decision mechanism and finally a cycle for a group of AAs that produces the action mechanism. The cycles are ordered in such a way that the result of a weaving cycle will be the base assembly for the next cycle of weaving. Thus, a component instantiated in a weaving cycle can be reused by AAs from the next weaving cycles through their pointcut and thus in their advices. This will allows a designer to divide an AA into several AAs. Then, AAs may be triggered in a cascaded way, i.e. the application of AAs for functionality from a concern in a cycle *n-1* may be the origin of the weaving of an AA in a cycle *n*. Thus, the cycle number 0 is always woven on an initial assembly blank of any AA. A weaving cycle *n* is always woven on the result of the weaving cycle *n-1*. A weaving cycle in this approach can be formally written as: $T(Ass_n; A_n) = Ass_{n+1}$ where $Ass_n$ is the assembly resulting from the weaving number **n**. The whole weaving process can be formally written as: $Ass_m = T^m(A_m, T^{m-1}(A_{m-1}, \ldots, T^0(A_0, Ass_0)))$.

The cascaded weaving of AA proceeds as follows: AAs for the first cycle are woven, on the resulting assembly, AAs for the second are woven and so on until the last cycle. Then, the whole process will be restarted, beginning with the cycle number 0. Each AA for functionality is woven with other AA for the same functionality. So between several AA for a same functionality (i.e. a same weaving cycle), the symmetry property of the weaving operation is preserved and interferences are managed.

Thanks to this decomposition, designing a concern will often consist in writing a combination of AAs, called a *Cascade of AAs*. All Cascades of AAs can be defined as follows: a Cascade of AAs is an ordered set of unordered sets of AA:

$$C = \{\{AA_{00}...AA_{0j}\}, ..., \{AA_{i0}...AA_{ij}\}\}$$

A Cascade of AAs can be decomposed as a set of cascades. The range of a set of AAs in a cascade defines the weaving cycle for which the set is designated. A Cascade of AAs does not necessarily contain a set of AAs for each cycle. Various Cascades of AAs for various concerns can be deployed simultaneously.

For example, in our scenario, we can identify two concerns and then two Cascades of AAs: (1) assistance to the person and (2) energy saving. The various AAs that we will present in this section are distributed as shown in Figure 7 in the various weaving cycles.

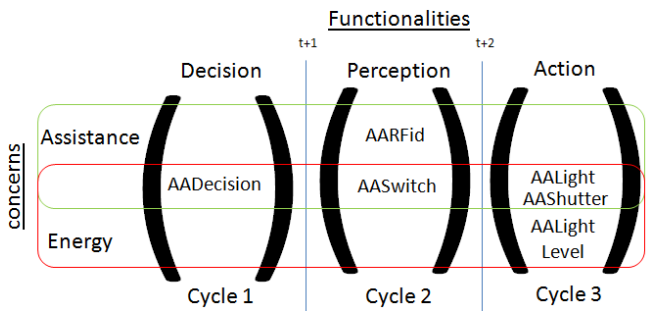

Fig. 7 Decision module (AADecision)

At first, we will describe the concerns of assistance to the person which has priority. This concern will involve three rounds of weaving. The Cascade of AAs designed for this concern is presented in Figure 7. Initially we will write a first AA (Fig. 8) for a first weaving cycle. This is the decision-making part of the system. It will be the link between the perception part and the action part of the system. Therefore, it will be heavily reused by other parts of the behavior. We could have deployed AAs for the perception mechanism first and AAs for decision in the second cycle so that the decision part would be deployed according to the perception mechanism. But, for this scenario, it would have meant rewriting many times the pointcuts part of the AADecision aspect according to the perception mechanisms required for its application.

AADecision aims to instantiate a timer and a component (decision) whose role is to indicate whether to turn the light on or to open the shutters according to an identifier and a time given as input.

```
Advice:
1 schema dec():
2 Decision : 'WComp.BasicBeans.DecisionEntity';
3 Timer : 'WComp.BasicBeans.Timer';
4 Average : 'WComp.BasicBeans.Average';
5
6 Timer.^Status _New_Evented_Value -> (Decision.SetTime)
```

Fig. 8 Decision module (AADecision)

In two AAs for a second weaving cycle, we describe the mechanism of perception that will be implemented in the application. These two AAs (Fig. 9) aim to connect the RFid reader and the switch to the decision component. So when a badge is read by the reader or when the switch changes its state, the decision-making module will be informed of it.

```
Pointcut:
1 RFid:=/rfid.*/
2 DecisionEntity:=/Decision[[:digit:]]/
Advice:
1 schema obs(DecisionEntity,RFid):
2 Rfid.^ value_Evented_NewValue->(DecisionEntity.Manage)

Pointcut:
1 switch:=/switch.*/
2 DecisionEntity:=/Decision[[:digit:]]/
Advice:
1 schema obs(DecisionEntity,switch):
2 switch.^ value_Evented_NewValue->(DecisionEntity.Manage)
```

Fig. 9 Perception modules for RFid and switch
(AARFid & AASwitch)

Finally, we must add some AAs (Fig. 10) to bind the decision part to lights and shutters. These AAs are destined to a third round of weaving. We design two AAs to ensure that the system is still running even in the absence of one of those two actuators.

```
Pointcut:
1 Shutter:=/Shutter.*/
2 DecisionEntity:=/Decision[[:digit:]]/
Advice:
1 schema action(DecisionEntity,Shutter):
2 DecisionEntity.^ ShutterManagementEvent->(Shutter.SetState)

Pointcut:
1 light:=/light [[:digit:]]/
2 DecisionEntity:=/Decision[[:digit:]]/
Advice:
1 schema ActionLight(light, DecisionEntity):
2 DecisionEntity.^ LightManagementEvent->(Light.SetState)
```

Fig. 10 Action modules for Store and Light
(AARollerShutter & AALight)

We will now consider the concern of energy consumption. Similarly this behavior can be decomposed. AAs for perception and AAs for decision from the other concern are reused. Finally we create an AA (Fig. 11) for the third weaving cycle to add a filter on a call to open the shutter and to redirect those calls to the lamp according to the brightness outside.

```
Pointcut:
1 lum:=/light[[:digit:]]/
2 Shutter:=/shutter[[:digit:]]/
3 Brightness:=/brightness.*/
Advice :
1 schema action(lum,Shutter,Brightness):
2 threshold : 'BasicBeans.Threshold' ( threshold = 10 )
3 Shutter.SetStatus->(
4           IF(threshold.reached){lum.setState}else{call}
5 )
6 Brightness.NewValue -> (Average.AValue)
7 Average.NewAverage -> (threshold.SetValue)
```

Fig. 11 Action module (AALightLevel)

Since the application of these Cascades of AAs is done at runtime, the reconfigurations of the system are also done at runtime according to the underlying software infrastructure. AAs from one cycle that are applied can collaborate, be composed, with AAs to be woven in next cycles dynamically. This composition is not explicit, meaning that an AA cannot embed a rule to trigger another AA.

Such compositions can be defined as opportunistic, since an AA from cycle for functionality is applied whenever it can. Since each AA is independent, each of them will be evaluated and implemented according to the underlying software infrastructure as classical AAs. Thus every AAs of each cycle can be applied independently. Achievable configurations of the systems are then numerous and performed at runtime as the composition of AAs. The multi-cycle approach improves the management of the variability of the system compared to the mono-cycle approach.

The number of achievable configuration for a set of AAs is calculable. When AAs from various cycles require in their pointcut, in order to be applied, some components from AAs previously woven, this number of configurations is reduced. In our scenario the aspect AADecision have to be applied in order to weave others AAs from cycle 2 and 3. In fact, such AAs can be considered as a single one, meaning that AADecision and AALight can be consider as a single AA. Then the number of types of configurations that can be achieved in the multi-cycle approach is described in Figure 12.

$$C = \prod_{i=0}^{K} 2^{M(i) - R(i)}$$

$I$: Cycle identifier
$K$: Number of cycle
$M(i)$ : Number of AAs involved in the cycle number $i$
$R(i)$ : Number of AAs involved in the cycle number $i$ that produce a black-box component that will be a requirement for future AAs to be woven
$C$: Number of combinations

Fig. 12 Number of configuration that can be generated

In the scenario, action and perception concerns of the system require, to be applied, the presence of the decision part. The number of type of configurations that can be achieved thanks to these cascades of AAs is $2^2 \times 2^3 = 32$. Using a mono-cycle weaving we could achieve $2^2=4$ configurations.

This ability to combine various AAs at runtime, more than increasing the number of reconfigurations that can be achieved using a minimal number of AAs, also serves to increase the adaptability of applications to their infrastructure for greater continuity of service, and greater variability. Indeed, the various functionalities associated to the various weaving cycles can be implemented in various ways, according to AAs that can be applied. During an appearance or disappearance of a device in the software infrastructure of the application, the AAs that can be applied are woven in an opportunistic way. The concern to be set up in a weaving cycle is then always implemented with the maximum AAs applied depending on the underlying infrastructure. In this way, the loss in the infrastructure of a device, used for a feature, does not lead necessarily to the loss of the feature in the application. Only parts of the functionality that cannot be woven are no longer implemented. Similarly it becomes possible to provide alternative mechanisms for these functionalities. So, if a device is available and can do the same as the one that just disappear, it can be used to replace it at runtime. It adds variability and self-adaptation facilities to the specific concerns addressed by a group of AAs. Moreover, it provides a mechanism to manage the unpredictability of ubiquitous systems.

As an example to change or add new sensors for location and identification, only some AAs, similar to those previously described (Fig. 9), need to be added. Several AAs can be deployed simultaneously based on various identification devices and can be applied indiscriminately. Thus, sometimes the system will work with all these sensors, sometimes only with some of them; and this without to have to worry about it, because it is done at runtime, once the AAs are deployed.

## 2.5 Multi-cycle merging

Various Cascades of AAs can also be composed. It consists in the union of the sets of AAs of the same range (i.e. to be woven in the same cycle). That is to say, AAs from various cascades to adapt a same functionality are all deployed in a same set.

$$C_a \cup C_b = \{\{AA_{00}...AA_{0j}\}_a \cup \{AA_{00}...AA_{0j}\}_b ,..., \{AA_{i0}...AA_{ij}\}_a \cup \{AA_{i0}...AA_{ij}\}_b\}$$

The union operator is symmetric, so the order in which combination are composed is not important. So the weaving operation of various Cascades of AAs is symmetric.

But the multi-cycle approach introduces a new type of interferences between several weaving cycles. An AA for a weaving cycle can have a side effect on AAs for next cycles. An AA for a concern may trigger an AA from a next cycle for another concern. This may be the cause of an adverse side effect on the reconfiguration of the system. The reverse is not possible. An aspect cannot remove a component that was required to weave another aspect. This type of interaction can be managed using namespace. To each cascade can be associated a name and a namespace. All the AAs included in the cascade, if they do not declare their own namespace, belong to the namespace of the cascade. An AA can declare its own namespace. Thus, two AAs with the same base name, but belonging to two cascades will not be the same if the two cascades do not share the same namespace. Thus an AA belonging to a namespace can reuse component from AAs from previous cycles that belong to the same namespace. To achieve this, to each component generated by an AA is associated the namespace of the AA. Interactions can be managed in three ways: (1) a cascade can be in a global namespace and thus all other AA from other cascades can interact with it; (2) the cascade is sharing its namespace with another cascade and thus only the cascades in the same namespace can interact one with each other; (3) the cascade do not share its namespace with others cascades, thus no interactions between cascades are permitted.

## 2.6 Synthesis

Aspects of Assembly are a mechanism to achieve **compositional adaptation** of components assemblies. The aspect oriented approach is pushed to its climax meaning that **everything is aspect**. The application is described by a set of aspects. The bootstrap is then the set of appearing and disappearing components. Aspects are triggered at **runtime** in response to changes in the operational context of an application or in user preferences in an **every time weaving process**. AAs are described using a constrained language. The weaving process can be mono or multi-cycle using some sets of set of aspects in what we call **Cascade of Aspects of Assembly**. The multi-cycle approach allows managing the high **variability** of the operational context of an application by **combining AAs** in an opportunistic and not explicit way. We can thus describe many configuration of an ambient system using few aspects.

The merging mechanism embedded in the AA's weaver **ensures the functional consistency** of the adapted application. Moreover, because the symmetry property of the weaving operation is guaranteed whatever the approach (mono or multi-cycle), it allows to define AAs or cascaded AAs as some **independent adaptation entities**. No explicit dependencies are defined between Aspects of Assembly. Thus, concerns can be implemented **without anticipating changes** in the context of the target application.

## 3. Experiments and validation

As mentioned earlier, response time of the adaptation process is a major concern in ubiquitous computing. It should be mastered and offer dynamics consistent with those of the changing environment. The frequency of adaptations that can be tolerated has to be as close as possible to the frequency of changes in the environment.

We evaluate our approach in term of performance with some experiments on the duration of a weaving cycle over components assemblies randomly generated. They were conducted on a standard personal laptop (Athlon X2, 1.6 GHz, 512Mo RAM). For this purpose various types of components have been instantiated randomly at runtime, in order to activate randomly two types of AAs. The number of joinpoints varies from 0 to 120 in these experiments and is directly related to the number of woven instance of advice.

### 3.1 Mono-cycle weaving duration

In term of duration, a weaving cycle can be divided into three major steps: (1) pointcut matching and combination, (2) merging and (3) translation of the resulting instance of advice into elementary instructions. During this time of adaptation, the weaver is no longer open to other disruptions; it doesn't consider anymore changes occurring in the software infrastructure or on the selection and unselection of AAs by the user. Steps (1) and (3) have a low cost in time, indeed the joinpoint model involves only few data and the order in which they are processed do not matter. During a weaving cycle, the merging process is the most expensive in time. However, several instances of advice are not necessarily conflicting. Therefore, the cost in time of the composition process can be described as in Figure 13 and is directly related to the cost of the merging operation and its probability as noted in [24].

Figure 14 shows the evolution of the duration of the composition process without conflicts (i.e. $p_i=0$) according to the number of joinpoint given as input. We can see that this process is not time consuming

$$F = a_1 \cdot g_0 \times \sum_{i=1}^{y} w_i \cdot p_i \cdot M + a_2$$

$F$: duration of instance of advice merging
$g_o$: number of rules in the base assembly
$y$: number of instance of advice
$w$: number of advice rule
$a_1, a_2$: model parameters
$p_i$: merging probability
$M$: Cost of merging

Fig. 13 Duration of instance of advice composition

In contrast, Figure 15 s hows the duration of the composition process when it involves the merging mechanism. In the first curve, $p_i=0,33$ whereas in the second curve $p_i=0,5$. When involving the merging engine, the composition process is much more time consuming and the number of conflicting rules is a key parameter.
During a weaving cycle, when the merging probability is about 33%, the duration of the composition process represents over 80% of the global duration of the weaving operation. The curve presented in Figure 24 shows the evolution of this duration.

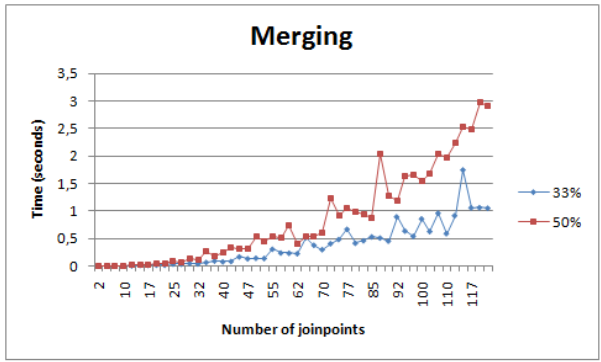

Fig.15 Duration of the composition process with $p_i=0,33$ and $p_i=0,5$

In the worst case, the composition operation involves rules that are all conflicting (where conflicts are all different). Thus, the cost of such a case describes the upper bound of the cost in time of the composition operation. It depends of the number of rules to be merged. An AA can be written as follows $AA_i=\{pointcut_i, advice_i\}$ where $advice_i=\{rule_{i0},$ … $rule_{ij}\}$. Thus, the number of rules to compose is the sum of all rules of all advices:

$$nbRule = \sum_{i=0}^{card(A_n)} card(advice_i)$$

Fig 16. Number of rules to be merged

Remind that the composition operation can be written as follows: $T(App_0, A_n)=App_1$ where $A_n=\{AA_0,…,AA_n\}$ where $App_0$ is the base application. $App_0$ is considered as a set of rules such as add components and bindings. Accordingly, the cost in time ($cT$) of the mono-cycle composition operation, in the worst case, can be expressed as follows:

$$cT = \left[\left(2^{nbRule} - (nbRule + 1)\right) \times card(App_0)\right] \times (cost\ to\ merge\ 2\ rules)$$

Fig. 17 Upper bound of the cost in time of the composition operation in the mono-cycle approach

Indeed, considering that the merging operation is symmetric and that all rules are conflicting (and that all conflicts are different from each other), the number of merging operation, between two rules, to be processed is: $(2^{nbRule} - (nbRule + 1))$.

The same goes for the pointcut matching process. This process aims to identify sets of joinpoints, a set for each pointcut rule. Then, it produces all possible combinations from these sets. A combination is a tuple including one joinpoint from each rule. This process is done independently for each aspect. The cost in time of the pointcut matching process in the mono-cycle weaving approach is the cost of the slowest process among all the AAs. In the worst case it depends on the number of combinations that must be calculated ($nbJPoint$ is the number of joinpoints):

$$nbJPoint^{card(pointcut_i)}$$

Fig. 18 Number of combination to be calculated in the mono-cycle approach

3.2 Multi-cycle weaving duration

In the multi-cycle approach the time spent to manage the chaining of cycles and the history of base assemblies is minimal. As we can see in Figure 20, this time is directly related to the number of cycles involved in the cascade. This figure presents the cost of the weaving process without composition and merging mechanisms. Thus, we can clearly see how the cost in time spent to manage cascades evolves according to the number of cycle.

As for the mono-cycle approach, in the worst case the composition operation involves, for each cycle, rules that are all conflicting with different conflicts. Therefore, the number of rules to compose in one cycle is:

$$nbRule_i = \sum_{j=0}^{card(A_i)} card(advice_j)$$

Fig. 19 Number of rules to be merged in one cycle

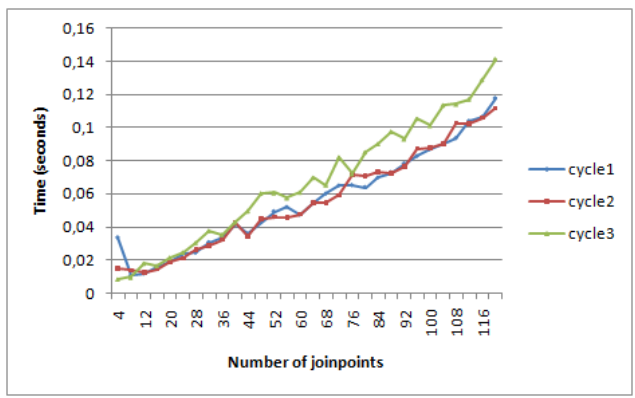

Fig. 20 Cost of the weaving process without the composition/merging engine

Remind that the weaving operation in the multi-cycle approach can be written as: $App_m = T^m(A_m, T^{m-1}(A_{m-1}, ..., T^0(A_0, App_0)))$. So, the cost in time of the composition operation ($cT^m$) in the multi-cycle approach can be expressed as follows:

$$cT^m = \left[\sum_{i=0}^{m}(2^{nbRule_i} - (nbRule_i + 1)) \times card(App_i)\right] \times (cost\ to\ merge\ 2\ rules)$$

Fig. 21 Upper bound of the cost in time of the composition operation in the multi-cycle approach

Indeed, considering that the merging operation is symmetric and that all rules are conflicting (and that all conflicts are different from each other), each rule is merged with all others for the same weaving cycle. So the number of merging operations to be processed is: $\sum_{i=0}^{m}(2^{nbRule_i} - (nbRule_i + 1))$

For the pointcut matching process we have seen that the cost of the operation depends on the number of combinations to be calculated. Using the multi-cycle approach, this number is:

$$\sum_{j=1}^{m} nbJPoint^{card(pointcut_j)}$$

Fig. 22 Number of configurations to be calculated in the multi-cycle approach.

### 3.3 Synthesis

We can see that, in order to implement a same functionality, depending on the chosen approach the costs of the composition operation may change. To compare both approaches, we consider that we can decompose a set of AAs as follows: $A_n = A_m \cup ... \cup A_0$. In the mono-cycle approach, all the sets $A_m, ..., A_0$ will be woven in a same cycle, whereas, in the multi-cycle, each set is woven in a different cycle. In such a case $cT^m \leq cT$ since both equation (Fig 21 and Fig 17) can be written as:

$$\left[\sum_{i=0}^{m}(2^{nbRule_i} - (nbRule_i + 1)) \times card(App_i)\right] \leq$$
$$\left[\prod_{i=0}^{m}(2^{nbRule_i} - (nbRule_i + 1))\right] \times card(App_0)$$

Fig. 23 Comparing Mono and multi-cycle composition cost in time

This is also true for the pointcut matching. Its cost in the mono-cycle approach is higher than its cost in the multi-cycle one:

$$\sum_{j=1}^{m} nbJPoint^{card(pointcut_j)}$$
$$\leq \prod_{j=1}^{m} nbJPoint^{card(pointcut_j)}$$

Fig. 24 Comparing Mono and multi-cycle pointcut matching cost

Thus, adaptation time, when using Aspects of Assembly or Cascaded Aspects of Assembly, is bounded by the adaptation time of the mono-cycle approach. When using AAs or Cascaded AAs, adaptation time is mastered and calculable. An important point is that decomposing an AA in order to use the multi-cycle approach, and then increasing the number of configurations described while designing few adaptation rules, is not a limiting factor with regard to the response time of the mechanism.

We can consider as standard, a set of adaptations schemas involving the merging mechanism in 33% of cases. In such a case the adaptation time can be modeled as in Figure 25.

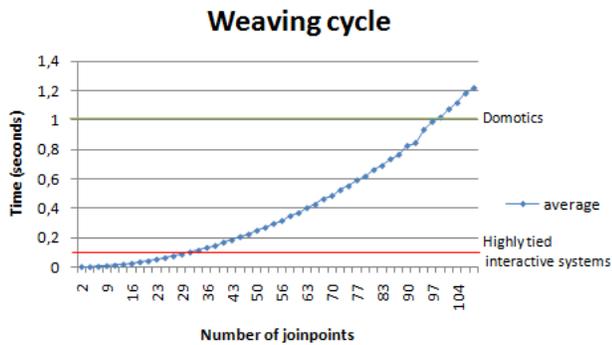

Fig. 25 Duration of weaving process

In the field of human computer interactions, it is considered that the user latency at most should be about 100ms. Then, Crowley *et al* in [26] propose that the latency for highly tied interactive systems must be twice lower than user latency: 50ms. Under this bound, we are able to compose about 30 joinpoints together in one cycle and about 10 AAs. On the other hand, ubiquitous computing does not necessarily require such a low response time. In the field of domotics, accepted latency is about 1 second.

As part of the Continuum project, our approach has been implemented, together with industrial partners, to represent an industrial scenario. This scenario takes place in the context of a hydrant man job. One of his is to close various valves in a water pipeline network, for the purposes of maintenance operations on the network. When undertaking the action of closing valves, our mobile worker is equipped either with a set of mobile devices, or with various devices in his car (GPS, Radar, Map, Camera, compass ...). The valves can also be equipped with various devices (humidity sensors, RFid ...). In this scenario, 18 AAs were written for 25 rules. In addition, between 7 and 10 devices are used, together with 7 off-the-shelf components for the user interface. The number of instances of advice generated thus ranges between 20 and 30, depending on the devices discovered. The number of interactions between identified adaptation rules ranges between 5 and 10, and such interactions appear in approximately 35% of cases. The response times observed and computed for the scenario are shorter than 50ms.

## 4. Related Works

Many works have identified the interest of aspects for ubiquitous or mobile computing, because of the encapsulation of adaptations into aspects [19,27]. For instance, in dynamic service adaptation [27], aspects are used to integrate services or to correct services mobile communication; they are not used to make structural reconfiguration of services orchestrations. Only few works allow achieving compositional adaptation and encapsulating adaptation into entities independent of each other. Moreover, amongst these works, only few propose adaptations with acceptable and mastered response time.

### 4.1. Logical properties

Aspects are not always independent of each other, some interactions may occur between them. In classical approaches, there is no support offered to resolve these interactions, it must be done explicitly by developers. The plugin architecture proposed in [28] is based on AO4BPEL [29] which is an aspect oriented workflow language. The latter allows dynamic adaptation of services compositions. In these works, the problem of management of interactions between aspects is not addressed dynamically. This management is implemented using the standard operators: *after, before...* Since this work is applied to workflows, they do not consider the dynamic evolution of the software infrastructure. In the proposed architecture there are two types of aspects: *monitoring aspects* that are able to activate or deactivate *adaptation aspects* at runtime. Aspects can be added, removed or sometimes generated at runtime. In our approach an AA and then combinations of AAs may also be added, removed or combined at runtime. An AA may also trigger of another AA. But in the case of AAs this is not necessarily defined explicitly (an AA does not describe in one rule that another AA may be triggered) for better reusability.

In EAOP [30], the authors propose mechanisms to define aspects of aspects. This mechanism allows applying aspects on others aspects including a mechanism to manage recursive calls. This is done using a monitor that sequentializes application of aspects. The monitor observes events from the execution of the application and spreads them to all aspects. The architecture is sequential, when the base application is stopped when it generates an event and involves the monitor. This is not the case with AA and Cascaded AAs. Moreover, AA's pointcuts do not concern the execution flow of the application but the structure of the component assembly to be applied.

JAsCo [31] is a dynamic AOP middleware. Aspects are encapsulated into components and connectors can deploy them by specifying their interactions. The aspects are woven according to a sequence of events represented as a finite state automaton. Advices can then be associated to the various transitions of this automaton. In this sense, aspects weaving can be chained. Like for the plugin architecture presented above, advices of the chain are well-defined and aspects are stateful which is not necessarily the case with the Cascaded AAs. On the other side, this approach allows to weave aspects according to the history of previously checked pointcuts.

Some works focus on the management or the detection of interferences between aspects. For example, Aksit *et al* [30] suggest a mechanism to identify interference issues and especially those on shared joinpoints. This approach is language independent. It consist in simulating and representing the various states of a program in the form of a graph and then identifying behavioral interferences between aspects, in particular with respect of the execution order of aspects. This type of approach for explicit resolution of interference issues can be found in many works [33,34,16]. In [16], many types of interferences are considered and addressed explicitly using policies.

As we already mentioned, this type of approach is hardly suitable within ubiquitous computing since we can not anticipate the adaptations that will be done on an application.

SAFRAN [2,35] is an extension of Fractal in order to facilitate the design of adaptive applications. To do this, they use adaptation aspects that can be added or removed at runtime. SAFRAN's joinpoint model is not restricted to the execution flow of the application. Adaptations can be triggered by some events related to the context of the application called exogenous events. The architecture of SAFRAN comprises two parts: (1) an adaptation language Fscript to reconfigure a component assembly where the ACID properties for dynamic reconfiguration are guaranteed; and (2) a toolkit to observe the context called WildCAT. An adaptation controller is integrated to the membrane to link, thanks to rules, these two parts and manage dependencies between adaptations explicitly. AA, conversely, don't require explicit dependencies, being independent from each other

4.2 Temporal properties

First of all, we have seen that because the environment is continuously evolving, the adaptation mechanism has to offer an every time adaptation process. Some works offer some adaptations that are not totally processed at runtime. In [36], Cheng *et al*. propose a mechanism to dynamically adapt applications that were not designed as adaptable. To achieve this, a two-stage process is implemented. The first is to implement, at design-time, some mechanisms that will thereafter allow the adaptation at runtime of an application. The second stage is to assess, at runtime, when to adapt and then to insert or remove some code in the application. Such two-step approach would be difficult to use in the field of ubiquitous computing because to implement adaptations some new unforeseen adaptations it would be necessary to go through step 1 again.

On the other hand, in most of current middleware for ubiquitous computing architectures, the software infrastructure is not specifically considered and is often subsumed in a global context [7,37,38]. For instance, SOCAM [7] is a middleware that allows rapid prototyping of context-aware services. SOCAM architecture offers a set of entities to automatically manage the perception and interpretation of the context including the software infrastructure. This often implies that the mechanism for context-awareness is based on an overall control loop. Thus, response time is often ignored by projects requiring complex context processing like ontologies, for which execution time is unbounded [6], sometimes requiring several seconds to process [7]. Consequently, response times are not mastered.

Conversely, some other approaches propose to decompose the context exploitation. In [38], Munelly *et al* propose to decompose the context into categories and to adapt an application according them using aspects. Aspects are used on top of classical objects. Such decomposition is interesting and allows considering several contexts separately. However, interferences between aspects are not managed and contextual information are in this approach some parameters of the adaptation. Unfortunately, aspects are triggered in a classical way and not according to changes occurring in the context.

## 5. Conclusion

We presented in this paper an approach for self-adaptation of ubiquitous applications. This approach allows reacting quickly with mastered response times, to changes occurring in the software infrastructure of the application to be adapted. Moreover, the merging mechanism implemented in the weaver ensures the independence of adaptations entities and the consistency of the resulting application. So, some adaptations can be designed and woven in an unforeseen way in order to build an application in an opportunistic way despite an unpredictable environment. Moreover, since these adaptations can be combined not explicitly thanks to a multi-cycle weaving process, the high variability of the software infrastructure can be managed with a minimum of adaptation rules. In future work, we will investigate whether it is possible to preprocess the whole or part of adaptation conflicts. To achieve this, the weaver should resolve as many conflicts as possible from abstract rules of advices. This would optimize the performance of the weaving process.

**Acknowledgments**

Thanks to Daniel Cheung-Foo-Wo for his early works on AA and evaluation of performances in his PhD Thesis. This work is supported by the French ANR Research Program VERSO in the project *ANR-08-VERS-005* called CONTINUUM.

**Nicolas Ferry** is preparing his PhD thesis on Context-aware and reactive adaptation of applications for pervasive computing at the University of Nice – Sophia Antipolis, supervised by Stéphane Lavirotte and Michel Riveill. His thesis is co-financed in a partnership between the CSTB (scientific and technical center for constructions) under the administrative supervision of the French Ministry of housing and the Provence Alpes Côte d'Azur Regional council.

**Jean-Yves Tigli** got his PhD degree in computer science from the University of Nice Sophia Antipolis, in 1996, on software engineering for intelligent robotics systems. He participated in various European projects between 1998 and 2002 (in ESPRIT and MAST European research programs). He's Associate Professor in Computer Science at the Engineering School of Technology of the University of Nice – Sophia Antipolis, France. He's currently managing and leading a project called "Continuum" supported by the French national research agency (ANR) to address the challenge of service continuity in dynamic pervasive environments involving various French universities and international companies.

**Stéphane Lavirotte** got his PhD degree in computer science from the University of Nice – Sophia Antipolis and INRIA, in 2000, on software for document Analysis and Recognition. He participated in various European projects between 1997 and 2004 (in ESPRIT, IST European research programs). He is Associate Professor in Computer Science at the IUFM of the University of Nice – Sophia Antipolis, France.

**Gaëtan Rey** got his PhD degree in computer science from the University of Joseph Fourrier at Grenoble, in 2005, on context-aware computing. During 2005-2006, he spent one year in the System Research Group of the University College of Dublin, UK. He's Associate Professor in Computer Science in the Institute of Technology of the University of Nice – Sophia Antipolis, France.

**Michel Riveill** got his PhD degree in computer science from the National Polytechnic Institute of Grenoble, in 1987, on distributed software. He obtained "Habilitation à Diriger les Recherches" in 1993, from the same institute. He's full Professor in Computer Science at the Engineering School of Technology of the University of Nice – Sophia Antipolis, France. He's leading the software engineering department of the computer science laboratory of the University of Nice - Sophia Antipolis and CNRS.